\documentclass[utf8]{frontiersSCNS} 

\usepackage{url,hyperref,lineno,microtype,subcaption}
\usepackage[onehalfspacing]{setspace}
\usepackage[flushleft]{threeparttable}
\usepackage[version=4]{mhchem}

\def\keyFont{\fontsize{8}{11}\helveticabold }
\def\firstAuthorLast{Bu\v{c}\'ik} 
\def\Authors{Radoslav Bu\v{c}\'ik\,$^{*}$}
\def\Address{Southwest Research Institute, San Antonio, TX, USA }
\def\corrAuthor{Radoslav Bu\v{c}\'ik}

\def\corrEmail{radoslav.bucik@swri.org}

\begin{document}
\onecolumn
\firstpage{1}

\title[Solar Energetic Particles]{Impulsive Solar Energetic Particle Events: EUV Waves and Jets } 

\author[\firstAuthorLast ]{\Authors} 
\address{} 
\correspondance{} 

\extraAuth{}

\maketitle

\begin{abstract}


Impulsive solar energetic particle (ISEP) events show peculiar elemental composition, with enhanced \ce{^{3}He} and heavy-ion abundances, markedly different from our solar system composition. Furthermore, the events are characterized by a wide variety of energy spectral shapes from power laws to rounded spectra toward the low energies. Solar sources of the events have been firmly associated with coronal jets. Surprisingly, new observations have shown that events are often accompanied by so-called extreme-ultraviolet (EUV) coronal waves -- a large-scale phenomenon compared to jets. This paper outlines the current understanding of the linkage of EUV waves with jets and energetic ions in ISEP events.       

\tiny
 \keyFont{ \section{Keywords:} solar energetic particles, element abundances, EUV waves, shock waves, solar jets, CMEs, flares} 
\end{abstract}

\section{Introduction}

Extreme-ultraviolet (EUV) waves appear as large-scale expanding disturbances in the corona. Since their discovery, there has been a long debate \citep[e.g.,][]{2012SoPh..281..187P,2014SoPh..289.3233L} whether they are true waves or just a magnetic field restructuring caused by associated coronal mass ejections (CMEs). Now, there is increasing evidence that they are true magnetosonic waves \citep[see][for a review]{2015LRSP...12....3W}. The EUV waves are regularly observed in CME-driven shock gradual solar energetic particle (GSEP) events 
\citep[e.g.,][]{1999ApJ...510..460T,2013ApJ...779..184P,2014ApJ...797....8L}, usually accompanied by large GOES X-ray (1--8\,{\AA}) flares (M-, X-class). 

Recently, with help of improved imaging resolution from Solar Terrestrial Relations Observatory (STEREO) and Solar Dynamics Observatory (SDO) that provide a full-disk view of the Sun from different observing angles, there have been reported EUV waves in many impulsive solar energetic particle (ISEP) events \citep{2013ApJ...762...54W,2015ApJ...806..235N,2015ApJ...812...53B,2016ApJ...833...63B,2021A&A...650A..23C}. Though these waves had a smaller spatial scale and were fainter than those in GSEP events, their observations in ISEP events, previously associated with compact flare signatures, were surprising. In this paper, we review the possible effects of the EUV waves on energetic ions in ISEP events and their relation to jets. 

\section{ISEP Events: Key Features}


ISEP events show tremendous (up to a factor of 10$^\text{4}$) enhancements of rare \ce{^{3}He} nuclide above the coronal composition \citep{1984SSRv...38...89K,2007SSRv..130..231M,2013SSRv..175...53R,2018SSRv..214...61R}. It is why these events are also called \ce{^{3}He}-rich. Heavy (\ce{^{20}Ne}--\ce{^{56}Fe}) and ultra-heavy ions (mass $>$70\,AMU; e.g., \ce{^{207}Pb}) are enhanced by a factor of 3--10 and $>$100, respectively, independently of the amount of \ce{^{3}He} enhancement \citep{1986ApJ...303..849M,1994ApJS...90..649R}. It has been interpreted as evidence that different mechanisms are involved in the acceleration of the \ce{^{3}He} and the heavy ions. In a typical ISEP event, the abundances of H, \ce{^{4}He}, \ce{^{12}C}, \ce{^{14}N}, \ce{^{16}O} are unenhanced. The enhancement of heavier ions increases monotonically with their mass. 

In addition to unusual abundances, these events show a rich variation of energy spectral shapes. Most events can be divided into two distinct groups: the class-1 events where all elements show similar power laws or broken power laws, and class-2 events with \ce{^{3}He} and Fe spectra curved toward low energies where H and intermediate species such as \ce{^{4}He} or O have spectra close to power law shape \citep{2000ApJ...545L.157M,2002ApJ...574.1039M,2016ApJ...823..138M,2015ApJ...806..235N,2016JPhCS.767a2002B}. It has been suggested that rounded spectra arise from a primary mechanism of \ce{^{3}He} (Fe) enrichment, and power laws involve a further stage of acceleration by a shock wave \citep{2000ApJ...545L.157M,2002ApJ...574.1039M}. Note, however, that type II radio bursts produced by shock acceleration of electrons in the corona are not characteristics of ISEP events. In contrast, these events show a high ($>$95\%) association with type III radio bursts \citep[e.g.,][]{1986ApJ...308..902R,2006ApJ...650..438N,2012ApJ...759...69W}, the emission generated by outward streaming $\sim$10--100 keV electron beams. The interplanetary propagation has not been thought to dominate the spectral shapes of \ce{^{3}He}-rich SEPs though for class-1 events with spectra of similar forms for all elements, this possibility cannot be ruled out \citep[e.g.,][]{2002ApJ...574.1039M}.

It is commonly accepted that unusual enrichments of \ce{^{3}He}-rich SEPs result from a unique acceleration mechanism associated with magnetic reconnection in solar flare sites. The models of \ce{^{3}He} acceleration involve ion-cyclotron resonance with plasma waves generated near the \ce{^{3}He} frequency \citep[e.g.,][]{1992ApJ...391L.105T,2006ApJ...636..462L}. Models of heavy-ion acceleration involve resonant interaction with cascading Alfv\'en waves \citep[e.g.,][]{1998SSRv...86...79M} or ion scattering on reconnecting magnetic islands \citep{2009ApJ...700L..16D}.

\section{ISEP Events: Jets}

\begin{figure}[h!]
\begin{center}
\includegraphics[width=0.48\textwidth,angle=270]{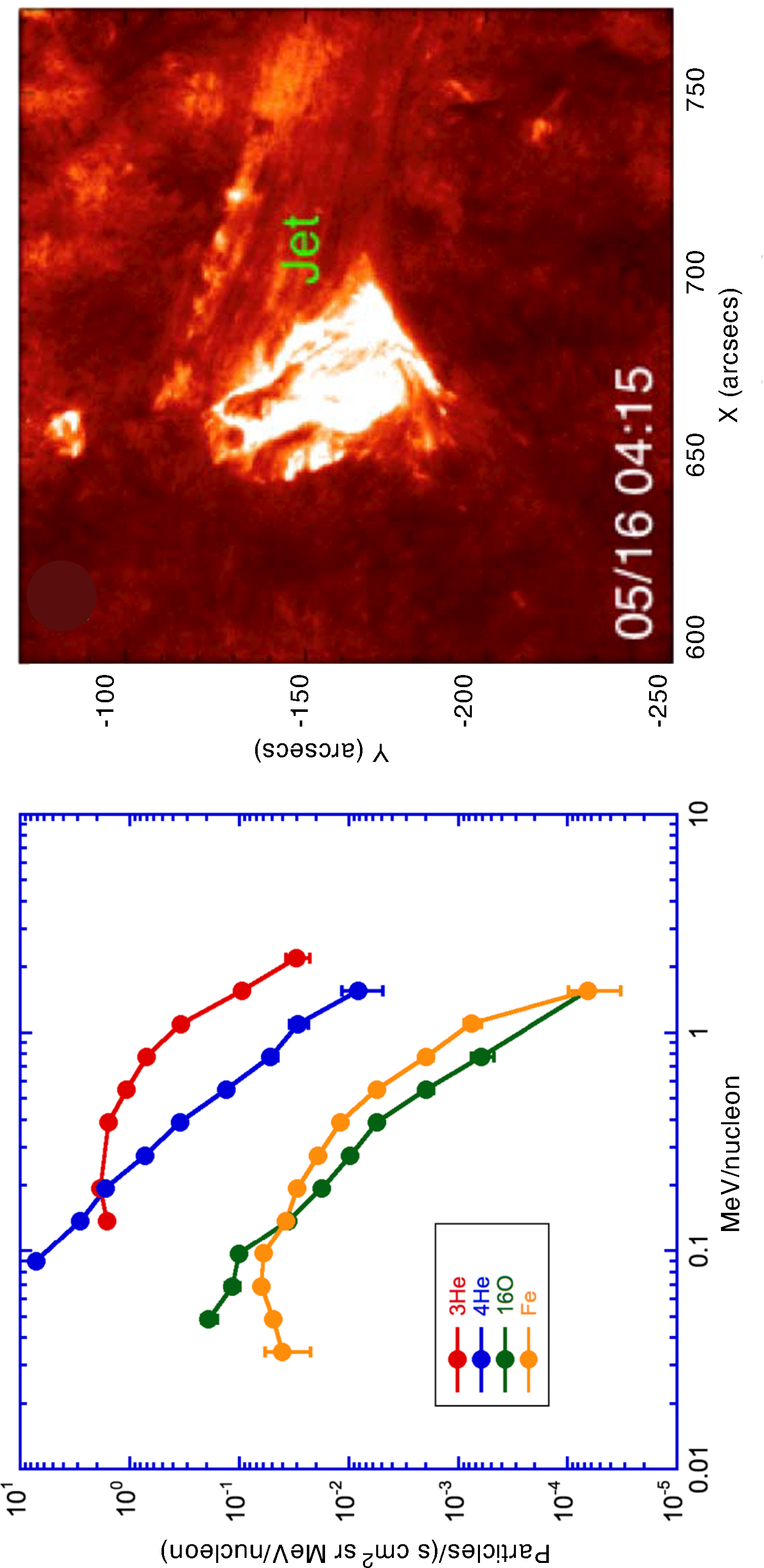}
\end{center}
\caption{The 2014 May 16 ISEP event. {\bf Left:} Advanced Composition Explorer (ACE) ULEIS energy spectra for selected ion species. The \ce{^{3}He} and Fe spectra are rounded, the \ce{^{4}He} and O show power law shape. Adapted from \citet{2015ApJ...806..235N}. \copyright AAS. Reproduced with permission. The \ce{^{3}He} /\ce{^{4}He}$\sim$14.9 over the energy range 0.5--2.0\,MeV/nucleon is extremely high. The ratio is a factor of $\sim$4$\times$10$^\text{4}$ higher than the solar wind value. The Fe/O at 0.385\,MeV/nucleon is $\sim$2.1. The \ce{^{28}Si} and \ce{^{32}S} had also curved spectra in this event \citep{2016ApJ...823..138M}. {\bf Right:} SDO AIA 304\,{\AA} direct image of the event solar source showing a jet eruption. Adapted from \citet{2016ApJ...823..138M}. \copyright AAS. Reproduced with permission. No X-ray flare was detected in this event.}\label{fig:1x}
\end{figure}

Source flares of \ce{^{3}He}-rich SEPs often show a jet-like shape in EUV and X-ray images \citep{2006ApJ...639..495W,2006ApJ...650..438N,2008ApJ...675L.125N,2015ApJ...806..235N,2020SSRv..216...24B,2020ApJS..246...42W} that is sometimes observed in coronagraphs as a narrow CME \citep{2001ApJ...562..558K,2006ApJ...639..495W,2012ApJ...759...69W}. Observation of jets in ISEP sources is believed to be a signature of ion acceleration via magnetic reconnection involving field lines open to interplanetary space \citep{2002ApJ...571L..63R,2006ApJ...639..495W}. Several events did not show jets but rather some amorphous brightening that has been attributed to instrument resolution and projection effects \citep{2006ApJ...639..495W}. The events are accompanied by small X-ray flares, typically B- or C-class. Figure~\ref{fig:1x} shows the energy spectra of a class-2 ISEP event and a jet that was the event source.

A concept of two basic populations of ISEP events have been suggested: SEP1 -- pure ISEP events from magnetic reconnection in solar jets and SEP2 -- a mix of SEP1 from jets and ambient population, reaccelerated by shocks driven by the same jets \citep{2020SSRv..216...20R,2021SSRv..217...72R,10.3389/fspas.2021.760261}. It was primarily motivated by measurements of enhanced H abundance in ISEP events with fast narrow CMEs \citep[e.g.,][]{2001ApJ...562..558K}. Recently, an excess in H has been reported for one ISEP event associated with a narrow (18$^{\circ}$) 450\,km/s CME \citep{2021A&A...650A..23C}. \citet{2018A&A...619A..34B} speculated that small-width shocks associated with fast and narrow CMEs can contribute to the generation of ISEP events. The authors reported 24 very narrow ($<$20$^{\circ}$) and fast (median 724\,km/s) CMEs where most (20) of them were associated with \ce{^{3}He}-rich SEPs. \citet{2012ApJ...759...69W}, in their extensive statistical study, reported a CME median speed of 496\,km/s with median width of 47$^{\circ}$ in 624 electron \ce{^{3}He}-rich SEP events. \citet{2016A&A...585A.119W} reported $\sim$600--1100\,km/s CMEs with width $<$60$^{\circ}$ in ten electron \ce{^{3}He}-rich events. These works show that fast jet-like CMEs are not uncommon in ISEP events. 

\section{ISEP Events: EUV waves}

\subsection{Rate of Occurrence} \label{sec:occur}

When the association of \ce{^{3}He}-rich SEP events with jets became relatively well established, the reported occurrence of the EUV waves in solar sources of \ce{^{3}He}-rich SEPs \citep{2013ApJ...762...54W,2015ApJ...806..235N,2015ApJ...812...53B,2016ApJ...833...63B} was surprising. More than half (20 of 32) of \ce{^{3}He}-rich events during solar minimum conditions in 2007--2010 were accompanied by EUV waves, and the remaining events were associated with jets or brightenings \citep{2016ApJ...833...63B}. Half of the 26 examined \ce{^{3}He}-rich events in 2010--2014 were accompanied by jets, four events by EUV waves, and the remaining events by an eruption showing larger angular expansion than a jet \citep{2015ApJ...806..235N}. A discrepancy in the number of the observed EUV waves can be due to the event selection criteria and subjectivity in the classification of the flare shape. \citet{2016ApJ...833...63B}  selected all events in the examined period, also, the events near the detection threshold, while \citet{2015ApJ...806..235N} selected events with a clear \ce{^{3}He} injection or with clear \ce{^{3}He} presence preceded by $>$40\,keV electron event. We note that the eruptions associated with the 2010 October 17 and 2012 May 14 ISEP events in the study of \citet{2015ApJ...806..235N} are marked by \citet{2016ApJ...833...63B} and \citet{2018ApJ...861..105S} as EUV waves, respectively. \citet{2021A&A...650A..23C} identified three ISEP events at $\sim$0.3--0.5\,au where one was associated with the jet and the other two with signatures of EUV waves.

\subsection{Delays}

The EUV waves were seen to propagate toward the spacecraft magnetic foot-point on the Sun, wherein more than half of wave fronts crossed the foot-point \citep{2016ApJ...833...63B}. The crossings were observed between 5 and 40 minutes after type III burst onset. Interestingly, ion delayed injections around 60 minutes after type III radio bursts were reported by \citet{2016A&A...585A.119W}. \citet{2003AdSpR..32.2679H} found ion delays of $>$40 min after $\sim$45\,keV electron release times in some \ce{^{3}He}-rich SEP events. These delays have been suggested to be due to particle scattering or the CME shock acceleration \citep{2003AdSpR..32.2679H,2016A&A...585A.119W}. The reported delays could also be related to the travel time of the EUV waves from the \ce{^{3}He}-rich SEP source to the spacecraft magnetic foot-point. However, no timing studies involving EUV waves and ion arrival times have been performed for ISEP events. The average uncertainties \citep[$\pm$45 min; e.g.,][]{2000ApJ...545L.157M} in the estimation of the $<$1\,MeV/nucleon ion release time on the Sun are higher than the reported travel time of the EUV wave from the source to the spacecraft magnetic foot-point.  

\subsection{Energy Spectra}

\begin{figure}[h!]
\begin{center}
\includegraphics[width=0.505\textwidth,angle=270]{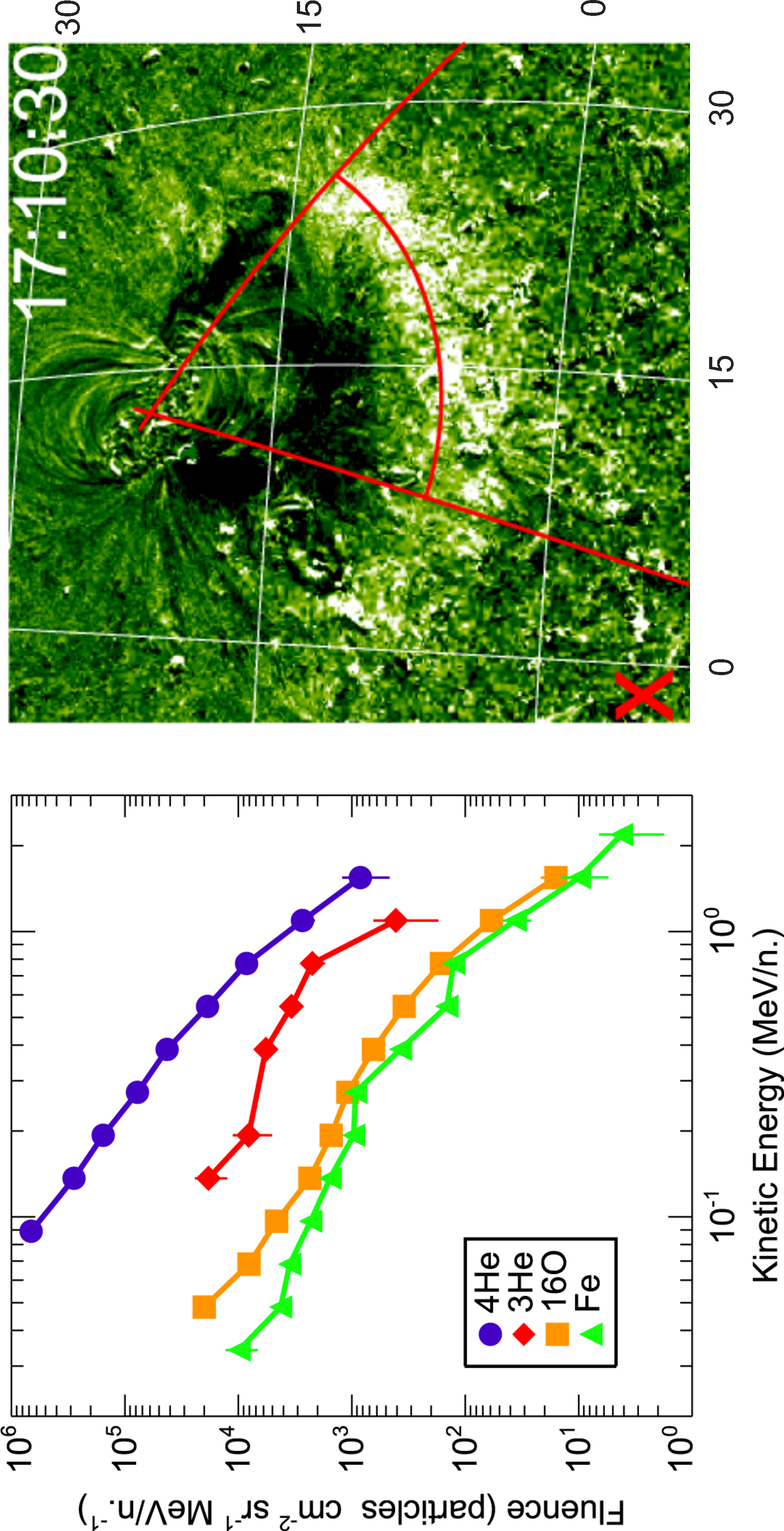}
\end{center}
\caption{The 2010 January 26 ISEP event. {\bf Left:} ACE ULEIS energy spectra for selected ion species. The \ce{^{4}He}, O, and Fe spectra show power law form; the \ce{^{3}He} spectrum is unambiguous. Adapted from \citet{2015ApJ...812...53B}. The \ce{^{3}He}/\ce{^{4}He} and Fe/O at 0.385\,MeV/nucleon are $\sim$0.13 and $\sim$0.6, respectively. The \ce{^{3}He}/\ce{^{4}He} is consistent with the median value of 0.385\,MeV/nucleon \ce{^{3}He}/\ce{^{4}He}$\sim$0.12 in class-1 events. The Fe/O is somewhat lower than the typical value in ISEP events \citep[$\sim$0.95 at 0.385\,MeV/nucleon;][]{2004ApJ...606..555M}. {\bf Right:} STEREO-A EUVI 195\,\AA \,5-min running difference image of the event solar source showing a coronal wave. Adapted from \citet{2015ApJ...812...53B}. Two red curves, passing through the source active region (AR), outline the propagation sector where the wave speed was determined. The red curve along the wave front outlines an arc centered on the AR. The red cross marks the ACE magnetic foot-point. The B-class X-ray flare was measured in the event.}\label{fig:2x}
\end{figure}

The wave kinematics has not been systematically investigated in ISEP events. The EUV wave expansion in two ISEP events was examined by \citet{2015ApJ...812...53B}. The authors reported the EUV wave front propagation speed of $\sim$300\,km/s which is comparable to the typical EUV wave speed \citep[$\sim$200--400\,km/s;][]{2009ApJS..183..225T}. The energy spectra in one event were typical of class-1 events where the associated wave showed a bright and sharp front. It has been suggested that the EUV waves with bright and sharp fronts may indicate shocks \citep{2002ApJ...569.1009B}. The energy spectra in the other event were typical of class-2 events where the wave showed a blurred and less bright front. Both these events also showed a jet in the source active region. The energy spectra for the ISEP event associated with a coronal wave are shown in Figure~\ref{fig:2x}~(Left). Figure~\ref{fig:2x}~(Right) displays the event-associated EUV wave, $\sim$8 minutes after type III radio burst or X-ray flare start time. The sharp wave front was seen later in the event. For four ISEP events reported in \citet{2016ApJ...833...63B}, the wave speed ($\sim$260--520\,km/s) was determined in earlier works \citep{2011A&A...532A.151W,2013ApJ...776...58N,2014SoPh..289.1257N}. All four events had class-1 energy spectra. The EUV wave speed of $\sim$600\,km/s for the 2011 January 27 ISEP event in \citet{2015ApJ...806..235N} was reported by \citet{2014SoPh..289.4563M}. The event had class-1 energy spectra. \citet{2013ApJ...776...58N} reported a EUV wave with a speed of $\sim$570\,km/s in the 2011 February 18 ISEP event \citep{2018ApJ...869L..21B} that was also associated with a jet. The event showed a double power law spectrum for \ce{^{3}He} at $\sim$0.1--15\,MeV/nucleon. The EUV wave speed in the ISEP event on 2012 May 14, mentioned in Section~\ref{sec:occur}, was $\sim$650\,km/s \citep{2018ApJ...861..105S}. The event showed class-1 spectra.

\begin{table}
\centering
\caption{\label{t1}The ISEP events with reported EUV wave speed.}
\vspace*{0.2cm}  
\begin{threeparttable}
{ \renewcommand{\arraystretch}{1.2}
\begin{tabular}{c l c c c c l}  
\hline\hline 
 &ISEP Event&Flare   &\ce{^{3}He}/\ce{^{4}He}&Fe/O&Wave Speed&References\\
 &Start Date  &Class  &                          &        &(km/s)          &                    \\
\hline
1&2007 May 23&B3.9&0.01\tnote{a}&2.30\tnote{a}&322&Event in 1; Speed from 2\\
2&2008 Nov 4	&C1.0&0.05\tnote{a}&1.45\tnote{a}&260&Event in 1; Speed from 2\\
3&2009 Dec 22&C7.2&1.55\tnote{a}&1.16\tnote{a}&403 (521)&Event in 1; Speed from 2 (3)\\
4&2010 Jan 26&B3.2&0.13\tnote{a}&0.60\tnote{a}&300&Event in 4; Speed from 4\\
5&2010 Feb 2&	…&0.46\tnote{a}&0.60\tnote{a}&$\gtrsim$200&Event in 4; Speed from 4\\
6&2010 Jun 12&M2.0&0.02\tnote{a}&0.94\tnote{a}&487 (386)&Event in 1; Speed from 3 (5)\\
7&2011 Jan 27&B6.6&0.08\tnote{b}&1.13\tnote{a}&610&Event in 6; Speed from 7\\
8&2011 Feb 18&C1.1&0.12\tnote{c}&1.46\tnote{c}&571&Event in 8; Speed from 5\\
9&2012 May 14&C2.5&0.05\tnote{b}&0.40\tnote{a}&648&Event in 6; Speed from 9\\
\hline
\end{tabular}
}
{\bf Notes.}
\begin{tablenotes}
        \item[a] 0.32--0.45\,MeV/nucleon
        \item[b] 0.50--2.00\,MeV/nucleon
        \item[c] 0.45--0.64\,MeV/nucleon\\
        {\bf References.} (1) \citet{2016ApJ...833...63B}, (2) \citet{2014SoPh..289.1257N}, (3) \citet{2011A&A...532A.151W}, (4) \citet{2015ApJ...812...53B}, (5) \citet{2013ApJ...776...58N}, (6) \citet{2015ApJ...806..235N}, (7) \citet{2014SoPh..289.4563M}, (8) \citet{2018ApJ...869L..21B}, (9)            \citet{2018ApJ...861..105S}
    \end{tablenotes}
    \end{threeparttable}
\end{table}

Table~\ref{t1} shows characteristics of the ISEP events with reported EUV wave speed. Column 1 indicates the event number, column 2 the ISEP event start date, column 3 GOES X-ray flare class. Columns 4 and 5 give \ce{^{3}He}/\ce{^{4}He} and Fe/O ratios, respectively. Column 6 provides the EUV wave speed. The last column indicates references where the events and speeds were reported. We can examine a correlation between elemental ratios, as shown in Table~\ref{t1}, and the EUV wave speed when more ISEP events with the estimated speed are available. Also, a correlation between the power law index and wave speed can be explored. It may help to understand the significance of EUV waves in ISEP events.

We see a tendency for ISEP events with jets only to have rounded \ce{^{3}He} and Fe spectra toward low energies and for events with coronal waves to have power law spectra \citep{2015ApJ...806..235N,2016JPhCS.767a2002B}. Thus, EUV waves may be related to some mechanism that modifies rounded spectra. Furthermore, \citet{2021A&A...650A..23C} have reported somewhat harder H and \ce{^{4}He} power law spectra in the two ISEP events with EUV wave than in the event with jet. Though the shock radio signature is not measured in ISEP events, it does not rule out that perhaps weak shocks that do not manifest as type II bursts can be formed in the flare site and re-accelerate \ce{^{3}He}-rich SEPs. It has been demonstrated with new observations \citep{2011A&A...532A.151W,2013ApJ...776...58N} that EUV waves at the early stages can be faster than the quiet-Sun fast magnetosonic speed \citep[$\sim$300\,km/s; e.g.,][]{2000ApJ...543L..89W}, implying that they can be shocks \citep{2011A&A...532A.151W}. 

\subsection{CMEs}

Not all ISEP events with EUV waves were found to be associated with CMEs. In some events only weak coronal outflows accompanied EUV waves. Generally, the associated CMEs were slow. Half of the ISEP events with EUV waves were accompanied by CMEs with a speed of $\lesssim$300\,km/s and a width of $\sim$50$^{\circ}$ \citep{2016ApJ...833...63B}. The ISEP events associated with EUV waves in the study of \citet{2015ApJ...806..235N} had a speed of $\sim$300\,km/s and a width of $\sim$50$^{\circ}$. One ISEP event with a EUV wave in the study of \citet{2021A&A...650A..23C} was accompanied by a CME with a speed of 450\,km/s and a width of 18$^{\circ}$. The other event was without a reported CME. 

\subsection{Jets}

In several ISEP events, EUV waves occurred together with jets \citep{2015ApJ...806..235N,2015ApJ...812...53B,2016ApJ...833...63B}; in other events only EUV waves were seen, possibly overwhelming the jet activity. An open question is the connection between jets and EUV waves in ISEP events and, ultimately, whether the energy spectra and the H abundance variations can be related to the jet-like CMEs or EUV waves. 

Only a few papers have examined the jet’s association with EUV waves. \citet{2012A&A...541A..49Z,2012ApJ...753..112Z} reported two EUV waves that were triggered by a jet. In one case the EUV wave was associated with a slow CME. \citet{2018ApJ...860L...8S} studied two EUV waves that were not associated with CME but were driven by loop expansion initiated by an accompanied jet. \citet{2018ApJ...861..105S} analyzed four recurrent jets, where each jet was accompanied by a narrow EUV wave ahead of the jet. In their study, only the last EUV wave was associated with (jet-like) CME. \citet{2018ApJ...869...39M} examined a EUV wave, associated with a slow CME, that appeared on top of the jet. These papers do not address an association with energetic particles. However, they show the EUV waves speed much higher than the corresponding jet speed. The authors pointed out that based on the properties (amplitude, speed, negative acceleration), these EUV waves should be regarded as fast magnetosonic waves or shocks.

\subsection{Longitude Spread}

Several authors discussed that the EUV waves in GSEP events may be linked with the injection of particles when the wave front intersects spacecraft magnetic foot-point 
\citep[e.g.,][]{1999ApJ...519..864K,1999ApJ...510..460T}, leading to a wide-longitude particle spread in the heliosphere \citep[e.g.,][]{2012ApJ...752...44R,2012AIPC.1436..259N,2013ApJ...779..184P,2015ApJ...808....3P,2014ApJ...797....8L,2014SoPh..289.3059R}. \citet{2014ApJ...797....8L} suggested that particles can also be accelerated at high altitudes without leaving EUV trace at the solar disk.

Interestingly, all nine ISEP events measured on widely separated ($\sim$40$^{\circ}$--80$^{\circ}$ in longitude) spacecraft \citep{2013ApJ...762...54W} were associated with EUV waves \citep{2015ApJ...806..235N,2016ApJ...833...63B}. In some cases, the waves were seen to cross the spacecraft magnetic foot-point \citep{2016ApJ...833...63B}. It has been discussed that the EUV waves may be also connected with widespread ISEP events \citep{2015ApJ...806..235N,2017ApJ...846..107Z}. The question is if EUV waves in small ISEP events directly accelerate particles or if they act indirectly, e.g., trigger the particle release by the expanding fronts \citep{1999ApJ...519..864K}.   

Other mechanisms for the wide longitude spread of \ce{^{3}He}-rich SEPs, in particular, cross-field diffusion and distortion of magnetic field lines by a CME, have been discussed elsewhere \citep[e.g.,][]{2013ApJ...762...54W}.

\section{Conclusion}

Previous works have suggested that the EUV waves in solar sources of ISEP events may be related to some features of these events, such as energy spectra variations, ion delays, and the wide longitude spread of energetic particles. The EUV wave speed was determined only in a few ISEP events with values ranging between $\sim$260 and 650\,km/s. The waves were observed together with jets or without jets, and they generally were accompanied by slow CMEs, or even CMEs were not observed. In some ISEP events, specifically in those without a CME, EUV waves were probably triggered by jets. 

New space missions at a close distance from the Sun, Parker Solar Probe, and Solar Orbiter may bring new insights on the role of EUV waves in ISEP events \citep[see][for the first reported ISEP events from these missions]{2020ApJS..246...42W,2021A&A...650A..23C,2020AAM,2021B}. For instance, measurements made close to the source of ISEP events remove uncertainties due to interplanetary transport effects.

\section*{Conflict of Interest Statement}

The author declares that the research was conducted in the absence of any commercial or financial relationships that could be construed as a potential conflict of interest.

\section*{Author Contributions}

All work on this manuscript was done by RB.

\section*{Funding}
RB was supported by NASA grant 80NSSC21K1316.

\section*{Acknowledgments}
The author thanks Nariaki Nitta for the critical review of the manuscript. This paper benefits from discussions within the International Space Science Institute (ISSI) Team ID 425 “Origins of \ce{^{3}He}-rich SEPs”.


\bibliographystyle{frontiersinSCNS_ENG_HUMS} 
\bibliography{fr}

\begin{thebibliography}{62}
\providecommand{\natexlab}[1]{#1}
\expandafter\ifx\csname urlstyle\endcsname\relax
  \providecommand{\doi}[1]{doi:\discretionary{}{}{}#1}\else
  \providecommand{\doi}{doi:\discretionary{}{}{}\begingroup
  \urlstyle{rm}\Url}\fi
\providecommand{\selectlanguage}[1]{\relax}
\providecommand{\bibAnnoteFile}[1]{%
  \IfFileExists{#1}{\begin{quotation}\noindent\textsc{Key:} #1\\
  \textsc{Annotation:}\ \input{#1}\end{quotation}}{}}
\providecommand{\bibAnnote}[2]{%
  \begin{quotation}\noindent\textsc{Key:} #1\\
  \textsc{Annotation:}\ #2\end{quotation}}

\bibitem[{{Biesecker} et~al.(2002){Biesecker}, {Myers}, {Thompson}, {Hammer},
  and {Vourlidas}}]{2002ApJ...569.1009B}
{Biesecker}, D.~A., {Myers}, D.~C., {Thompson}, B.~J., {Hammer}, D.~M., and
  {Vourlidas}, A. (2002).
\newblock {Solar Phenomena Associated with ``EIT Waves''}.
\newblock \emph{Astrophys. J.} 569, 1009--1015.
\newblock \doi{10.1086/339402}
\bibAnnoteFile{2002ApJ...569.1009B}

\bibitem[{{Bronarska} et~al.(2018){Bronarska}, {Wheatland}, {Gopalswamy}, and
  {Michalek}}]{2018A&A...619A..34B}
{Bronarska}, K., {Wheatland}, M.~S., {Gopalswamy}, N., and {Michalek}, G.
  (2018).
\newblock {Very narrow coronal mass ejections producing solar energetic
  particles}.
\newblock \emph{Astron. Astrophys.} 619, A34.
\newblock \doi{10.1051/0004-6361/201833237}
\bibAnnoteFile{2018A&A...619A..34B}

\bibitem[{{Bu{\v{c}}{\'\i}k}(2020)}]{2020SSRv..216...24B}
{Bu{\v{c}}{\'\i}k}, R. (2020).
\newblock {$^{3}$He-Rich Solar Energetic Particles: Solar Sources}.
\newblock \emph{Space Sci. Rev.} 216, 24.
\newblock \doi{10.1007/s11214-020-00650-5}
\bibAnnoteFile{2020SSRv..216...24B}

\bibitem[{{Bu{\v{c}}{\'\i}k} et~al.(2015){Bu{\v{c}}{\'\i}k}, {Innes}, {Guo},
  {Mason}, and {Wiedenbeck}}]{2015ApJ...812...53B}
{Bu{\v{c}}{\'\i}k}, R., {Innes}, D.~E., {Guo}, L., {Mason}, G.~M., and
  {Wiedenbeck}, M.~E. (2015).
\newblock {Observations of EUV Waves in $^{3}$He-rich Solar Energetic Particle
  Events}.
\newblock \emph{Astrophys. J.} 812, 53.
\newblock \doi{10.1088/0004-637X/812/1/53}
\bibAnnoteFile{2015ApJ...812...53B}

\bibitem[{{Bu{\v{c}}{\'\i}k} et~al.(2016{\natexlab{a}}){Bu{\v{c}}{\'\i}k},
  {Innes}, {Mason}, and {Wiedenbeck}}]{2016ApJ...833...63B}
{Bu{\v{c}}{\'\i}k}, R., {Innes}, D.~E., {Mason}, G.~M., and {Wiedenbeck}, M.~E.
  (2016{\natexlab{a}}).
\newblock {Association of $^{3}$He-Rich Solar Energetic Particles with
  Large-scale Coronal Waves}.
\newblock \emph{Astrophys. J.} 833, 63.
\newblock \doi{10.3847/1538-4357/833/1/63}
\bibAnnoteFile{2016ApJ...833...63B}

\bibitem[{{Bu{\v{c}}{\'\i}k} et~al.(2016{\natexlab{b}}){Bu{\v{c}}{\'\i}k},
  {Innes}, {Mason}, and {Wiedenbeck}}]{2016JPhCS.767a2002B}
{Bu{\v{c}}{\'\i}k}, R., {Innes}, D.~E., {Mason}, G.~M., and {Wiedenbeck}, M.~E.
  (2016{\natexlab{b}}).
\newblock {Energy spectra of $^{3}$He-rich solar energetic particles associated
  with coronal waves}.
\newblock \emph{J. Phys. Conf. Ser.} 767, 012002.
\newblock \doi{10.1088/1742-6596/767/1/012002}
\bibAnnoteFile{2016JPhCS.767a2002B}

\bibitem[{{Bu{\v{c}}{\'\i}k} et~al.(2021){Bu{\v{c}}{\'\i}k}, {Mason},
  {G\'omez-Herrero}, {Lario}, {Balmaceda}, {Nitta} et~al.}]{2021B}
{Bu{\v{c}}{\'\i}k}, R., {Mason}, G.~M., {G\'omez-Herrero}, R., {Lario}, D.,
  {Balmaceda}, L., {Nitta}, N.~V., et~al. (2021).
\newblock {The Long Period of $^{3}$He-rich Solar Energetic Particles Measured
  by Solar Orbiter on 2020 November 17-23}.
\newblock \emph{Astron. Astrophys.} 656, L11.
\newblock \doi{10.1051/0004-6361/202141009}
\bibAnnoteFile{2021B}

\bibitem[{{Bu{\v{c}}{\'\i}k} et~al.(2018){Bu{\v{c}}{\'\i}k}, {Wiedenbeck},
  {Mason}, {G{\'o}mez-Herrero}, {Nitta}, and {Wang}}]{2018ApJ...869L..21B}
{Bu{\v{c}}{\'\i}k}, R., {Wiedenbeck}, M.~E., {Mason}, G.~M.,
  {G{\'o}mez-Herrero}, R., {Nitta}, N.~V., and {Wang}, L. (2018).
\newblock {$^{3}$He-rich Solar Energetic Particles from Sunspot Jets}.
\newblock \emph{Astrophys. J. Lett.} 869, L21.
\newblock \doi{10.3847/2041-8213/aaf37f}
\bibAnnoteFile{2018ApJ...869L..21B}

\bibitem[{{Cohen} et~al.(2021){Cohen}, {Christian}, {Cummings}, {Davis},
  {Desai}, {de Nolfo} et~al.}]{2021A&A...650A..23C}
{Cohen}, C.~M.~S., {Christian}, E.~R., {Cummings}, A.~C., {Davis}, A.~J.,
  {Desai}, M.~I., {de Nolfo}, G.~A., et~al. (2021).
\newblock {Parker Solar Probe observations of He/H abundance variations in SEP
  events inside 0.5 au}.
\newblock \emph{Astron. Astrophys.} 650, A23.
\newblock \doi{10.1051/0004-6361/202039299}
\bibAnnoteFile{2021A&A...650A..23C}

\bibitem[{{Drake} et~al.(2009){Drake}, {Cassak}, {Shay}, {Swisdak}, and
  {Quataert}}]{2009ApJ...700L..16D}
{Drake}, J.~F., {Cassak}, P.~A., {Shay}, M.~A., {Swisdak}, M., and {Quataert},
  E. (2009).
\newblock {A Magnetic Reconnection Mechanism for Ion Acceleration and Abundance
  Enhancements in Impulsive Flares}.
\newblock \emph{Astrophys. J. Lett.} 700, L16--L20.
\newblock \doi{10.1088/0004-637X/700/1/L16}
\bibAnnoteFile{2009ApJ...700L..16D}

\bibitem[{{Ho} et~al.(2003){Ho}, {Roelof}, {Mason}, {Lario}, and
  {Mazur}}]{2003AdSpR..32.2679H}
{Ho}, G.~C., {Roelof}, E.~C., {Mason}, G.~M., {Lario}, D., and {Mazur}, J.~E.
  (2003).
\newblock {Onset study of impulsive solar energetic particle events}.
\newblock \emph{Advances in Space Research} 32, 2679--2684.
\newblock \doi{10.1016/S0273-1177(03)00930-X}
\bibAnnoteFile{2003AdSpR..32.2679H}

\bibitem[{{Kahler} et~al.(2001){Kahler}, {Reames}, and
  {Sheeley}}]{2001ApJ...562..558K}
{Kahler}, S.~W., {Reames}, D.~V., and {Sheeley}, J., N.~R. (2001).
\newblock {Coronal Mass Ejections Associated with Impulsive Solar Energetic
  Particle Events}.
\newblock \emph{Astrophys. J.} 562, 558--565.
\newblock \doi{10.1086/323847}
\bibAnnoteFile{2001ApJ...562..558K}

\bibitem[{{Kocharov} and {Kocharov}(1984)}]{1984SSRv...38...89K}
{Kocharov}, L.~G. and {Kocharov}, G.~E. (1984).
\newblock {$^{3}$He-rich solar flares}.
\newblock \emph{Space Sci. Rev.} 38, 89--141.
\newblock \doi{10.1007/BF00180337}
\bibAnnoteFile{1984SSRv...38...89K}

\bibitem[{{Krucker} et~al.(1999){Krucker}, {Larson}, {Lin}, and
  {Thompson}}]{1999ApJ...519..864K}
{Krucker}, S., {Larson}, D.~E., {Lin}, R.~P., and {Thompson}, B.~J. (1999).
\newblock {On the Origin of Impulsive Electron Events Observed at 1 AU}.
\newblock \emph{Astrophys. J.} 519, 864--875.
\newblock \doi{10.1086/307415}
\bibAnnoteFile{1999ApJ...519..864K}

\bibitem[{{Lario} et~al.(2014){Lario}, {Raouafi}, {Kwon}, {Zhang},
  {G{\'o}mez-Herrero}, {Dresing} et~al.}]{2014ApJ...797....8L}
{Lario}, D., {Raouafi}, N.~E., {Kwon}, R.~Y., {Zhang}, J., {G{\'o}mez-Herrero},
  R., {Dresing}, N., et~al. (2014).
\newblock {The Solar Energetic Particle Event on 2013 April 11: An
  Investigation of its Solar Origin and Longitudinal Spread}.
\newblock \emph{Astrophys. J.} 797, 8.
\newblock \doi{10.1088/0004-637X/797/1/8}
\bibAnnoteFile{2014ApJ...797....8L}

\bibitem[{{Liu} et~al.(2006){Liu}, {Petrosian}, and
  {Mason}}]{2006ApJ...636..462L}
{Liu}, S., {Petrosian}, V., and {Mason}, G.~M. (2006).
\newblock {Stochastic Acceleration of $^{3}$He and $^{4}$He in Solar Flares by
  Parallel-propagating Plasma Waves: General Results}.
\newblock \emph{Astrophys. J.} 636, 462--474.
\newblock \doi{10.1086/497883}
\bibAnnoteFile{2006ApJ...636..462L}

\bibitem[{{Liu} and {Ofman}(2014)}]{2014SoPh..289.3233L}
{Liu}, W. and {Ofman}, L. (2014).
\newblock {Advances in Observing Various Coronal EUV Waves in the SDO Era and
  Their Seismological Applications (Invited Review)}.
\newblock \emph{Sol. Phys.} 289, 3233--3277.
\newblock \doi{10.1007/s11207-014-0528-4}
\bibAnnoteFile{2014SoPh..289.3233L}

\bibitem[{{Mason}(2007)}]{2007SSRv..130..231M}
{Mason}, G.~M. (2007).
\newblock {$^{3}$He-Rich Solar Energetic Particle Events}.
\newblock \emph{Space Sci. Rev.} 130, 231--242.
\newblock \doi{10.1007/s11214-007-9156-8}
\bibAnnoteFile{2007SSRv..130..231M}

\bibitem[{{Mason} et~al.(2000){Mason}, {Dwyer}, and
  {Mazur}}]{2000ApJ...545L.157M}
{Mason}, G.~M., {Dwyer}, J.~R., and {Mazur}, J.~E. (2000).
\newblock {New Properties of $^{3}$He-rich Solar Flares Deduced from Low-Energy
  Particle Spectra}.
\newblock \emph{Astrophys. J. Lett.} 545, L157--L160.
\newblock \doi{10.1086/317886}
\bibAnnoteFile{2000ApJ...545L.157M}

\bibitem[{{Mason} et~al.(2021){Mason}, {Ho}, {Allen}, {Rodr{\'\i}guez-Pacheco},
  {Wimmer-Schweingruber}, {Bu{\v{c}}{\'\i}k} et~al.}]{2020AAM}
{Mason}, G.~M., {Ho}, G.~C., {Allen}, R., {Rodr{\'\i}guez-Pacheco}, J.,
  {Wimmer-Schweingruber}, R.~F., {Bu{\v{c}}{\'\i}k}, R., et~al. (2021).
\newblock {$^{3}$He-rich solar energetic particle events observed on the first
  perihelion pass of Solar Orbiter}.
\newblock \emph{Astron. Astrophys.} 656, L1.
\newblock \doi{10.1051/0004-6361/202039752}
\bibAnnoteFile{2020AAM}

\bibitem[{{Mason} et~al.(2004){Mason}, {Mazur}, {Dwyer}, {Jokipii}, {Gold}, and
  {Krimigis}}]{2004ApJ...606..555M}
{Mason}, G.~M., {Mazur}, J.~E., {Dwyer}, J.~R., {Jokipii}, J.~R., {Gold},
  R.~E., and {Krimigis}, S.~M. (2004).
\newblock {Abundances of Heavy and Ultraheavy Ions in $^{3}$He-rich Solar
  Flares}.
\newblock \emph{Astrophys. J.} 606, 555--564.
\newblock \doi{10.1086/382864}
\bibAnnoteFile{2004ApJ...606..555M}

\bibitem[{{Mason} et~al.(2016){Mason}, {Nitta}, {Wiedenbeck}, and
  {Innes}}]{2016ApJ...823..138M}
{Mason}, G.~M., {Nitta}, N.~V., {Wiedenbeck}, M.~E., and {Innes}, D.~E. (2016).
\newblock {Evidence for a Common Acceleration Mechanism for Enrichments of
  $^{3}$He and Heavy Ions in Impulsive SEP Events}.
\newblock \emph{Astrophys. J.} 823, 138.
\newblock \doi{10.3847/0004-637X/823/2/138}
\bibAnnoteFile{2016ApJ...823..138M}

\bibitem[{{Mason} et~al.(1986){Mason}, {Reames}, {Klecker}, {Hovestadt}, and
  {von Rosenvinge}}]{1986ApJ...303..849M}
{Mason}, G.~M., {Reames}, D.~V., {Klecker}, B., {Hovestadt}, D., and {von
  Rosenvinge}, T.~T. (1986).
\newblock {The Heavy-Ion Compositional Signature in $^{3}$He-rich Solar
  Particle Events}.
\newblock \emph{Astrophys. J.} 303, 849.
\newblock \doi{10.1086/164133}
\bibAnnoteFile{1986ApJ...303..849M}

\bibitem[{{Mason} et~al.(2002){Mason}, {Wiedenbeck}, {Miller}, {Mazur},
  {Christian}, {Cohen} et~al.}]{2002ApJ...574.1039M}
{Mason}, G.~M., {Wiedenbeck}, M.~E., {Miller}, J.~A., {Mazur}, J.~E.,
  {Christian}, E.~R., {Cohen}, C.~M.~S., et~al. (2002).
\newblock {Spectral Properties of He and Heavy Ions in $^{3}$He-rich Solar
  Flares}.
\newblock \emph{Astrophys. J.} 574, 1039--1058.
\newblock \doi{10.1086/341112}
\bibAnnoteFile{2002ApJ...574.1039M}

\bibitem[{{Miao} et~al.(2018){Miao}, {Liu}, {Li}, {Shen}, {Yang}, {Elmhamdi}
  et~al.}]{2018ApJ...869...39M}
{Miao}, Y., {Liu}, Y., {Li}, H.~B., {Shen}, Y., {Yang}, S., {Elmhamdi}, A.,
  et~al. (2018).
\newblock {A Blowout Jet Associated with One Obvious Extreme-ultraviolet Wave
  and One Complicated Coronal Mass Ejection Event}.
\newblock \emph{Astrophys. J.} 869, 39.
\newblock \doi{10.3847/1538-4357/aaeac1}
\bibAnnoteFile{2018ApJ...869...39M}

\bibitem[{{Miller}(1998)}]{1998SSRv...86...79M}
{Miller}, J.~A. (1998).
\newblock {Particle Acceleration in Impulsive Solar Flares}.
\newblock \emph{Space Sci. Rev.} 86, 79--105.
\newblock \doi{10.1023/A:1005066209536}
\bibAnnoteFile{1998SSRv...86...79M}

\bibitem[{{Muhr} et~al.(2014){Muhr}, {Veronig}, {Kienreich}, {Vr{\v{s}}nak},
  {Temmer}, and {Bein}}]{2014SoPh..289.4563M}
{Muhr}, N., {Veronig}, A.~M., {Kienreich}, I.~W., {Vr{\v{s}}nak}, B., {Temmer},
  M., and {Bein}, B.~M. (2014).
\newblock {Statistical Analysis of Large-Scale EUV Waves Observed by
  STEREO/EUVI}.
\newblock \emph{Sol. Phys.} 289, 4563--4588.
\newblock \doi{10.1007/s11207-014-0594-7}
\bibAnnoteFile{2014SoPh..289.4563M}

\bibitem[{{Nitta}(2012)}]{2012AIPC.1436..259N}
{Nitta}, N.~V. (2012).
\newblock {Magnetic field connection and large scale coronal disturbances in
  the context of gradual SEP events}.
\newblock In \emph{Physics of the Heliosphere: A 10 Year Retrospective}, eds.
  J.~{Heerikhuisen}, G.~{Li}, N.~{Pogorelov}, and G.~{Zank}. vol. 1436 of
  \emph{American Institute of Physics Conference Series}, 259--264.
\newblock \doi{10.1063/1.4723617}
\bibAnnoteFile{2012AIPC.1436..259N}

\bibitem[{{Nitta} et~al.(2014){Nitta}, {Aschwanden}, {Freeland}, {Lemen},
  {W{\"u}lser}, and {Zarro}}]{2014SoPh..289.1257N}
{Nitta}, N.~V., {Aschwanden}, M.~J., {Freeland}, S.~L., {Lemen}, J.~R.,
  {W{\"u}lser}, J.~P., and {Zarro}, D.~M. (2014).
\newblock {The Association of Solar Flares with Coronal Mass Ejections During
  the Extended Solar Minimum}.
\newblock \emph{Sol. Phys.} 289, 1257--1277.
\newblock \doi{10.1007/s11207-013-0388-3}
\bibAnnoteFile{2014SoPh..289.1257N}

\bibitem[{{Nitta} et~al.(2015){Nitta}, {Mason}, {Wang}, {Cohen}, and
  {Wiedenbeck}}]{2015ApJ...806..235N}
{Nitta}, N.~V., {Mason}, G.~M., {Wang}, L., {Cohen}, C. M.~S., and
  {Wiedenbeck}, M.~E. (2015).
\newblock {Solar Sources of $^{3}$He-rich Solar Energetic Particle Events in
  Solar Cycle 24}.
\newblock \emph{Astrophys. J.} 806, 235.
\newblock \doi{10.1088/0004-637X/806/2/235}
\bibAnnoteFile{2015ApJ...806..235N}

\bibitem[{{Nitta} et~al.(2008){Nitta}, {Mason}, {Wiedenbeck}, {Cohen},
  {Krucker}, {Hannah} et~al.}]{2008ApJ...675L.125N}
{Nitta}, N.~V., {Mason}, G.~M., {Wiedenbeck}, M.~E., {Cohen}, C. M.~S.,
  {Krucker}, S., {Hannah}, I.~G., et~al. (2008).
\newblock {Coronal Jet Observed by Hinode as the Source of a$^{3}$He-rich Solar
  Energetic Particle Event}.
\newblock \emph{Astrophys. J. Lett.} 675, L125.
\newblock \doi{10.1086/533438}
\bibAnnoteFile{2008ApJ...675L.125N}

\bibitem[{{Nitta} et~al.(2006){Nitta}, {Reames}, {De Rosa}, {Liu}, {Yashiro},
  and {Gopalswamy}}]{2006ApJ...650..438N}
{Nitta}, N.~V., {Reames}, D.~V., {De Rosa}, M.~L., {Liu}, Y., {Yashiro}, S.,
  and {Gopalswamy}, N. (2006).
\newblock {Solar Sources of Impulsive Solar Energetic Particle Events and Their
  Magnetic Field Connection to the Earth}.
\newblock \emph{Astrophys. J.} 650, 438--450.
\newblock \doi{10.1086/507442}
\bibAnnoteFile{2006ApJ...650..438N}

\bibitem[{{Nitta} et~al.(2013){Nitta}, {Schrijver}, {Title}, and
  {Liu}}]{2013ApJ...776...58N}
{Nitta}, N.~V., {Schrijver}, C.~J., {Title}, A.~M., and {Liu}, W. (2013).
\newblock {Large-scale Coronal Propagating Fronts in Solar Eruptions as
  Observed by the Atmospheric Imaging Assembly on Board the Solar Dynamics
  Observatory{\textemdash}an Ensemble Study}.
\newblock \emph{Astrophys. J.} 776, 58.
\newblock \doi{10.1088/0004-637X/776/1/58}
\bibAnnoteFile{2013ApJ...776...58N}

\bibitem[{{Park} et~al.(2013){Park}, {Innes}, {Bucik}, and
  {Moon}}]{2013ApJ...779..184P}
{Park}, J., {Innes}, D.~E., {Bucik}, R., and {Moon}, Y.~J. (2013).
\newblock {The Source Regions of Solar Energetic Particles Detected by Widely
  Separated Spacecraft}.
\newblock \emph{Astrophys. J.} 779, 184.
\newblock \doi{10.1088/0004-637X/779/2/184}
\bibAnnoteFile{2013ApJ...779..184P}

\bibitem[{{Park} et~al.(2015){Park}, {Innes}, {Bucik}, {Moon}, and
  {Kahler}}]{2015ApJ...808....3P}
{Park}, J., {Innes}, D.~E., {Bucik}, R., {Moon}, Y.~J., and {Kahler}, S.~W.
  (2015).
\newblock {Study of Solar Energetic Particle Associations with Coronal
  Extreme-ultraviolet Waves}.
\newblock \emph{Astrophys. J.} 808, 3.
\newblock \doi{10.1088/0004-637X/808/1/3}
\bibAnnoteFile{2015ApJ...808....3P}

\bibitem[{{Patsourakos} and {Vourlidas}(2012)}]{2012SoPh..281..187P}
{Patsourakos}, S. and {Vourlidas}, A. (2012).
\newblock {On the Nature and Genesis of EUV Waves: A Synthesis of Observations
  from SOHO, STEREO, SDO, and Hinode (Invited Review)}.
\newblock \emph{Sol. Phys.} 281, 187--222.
\newblock \doi{10.1007/s11207-012-9988-6}
\bibAnnoteFile{2012SoPh..281..187P}

\bibitem[{{Reames}(2002)}]{2002ApJ...571L..63R}
{Reames}, D.~V. (2002).
\newblock {Magnetic Topology of Impulsive and Gradual Solar Energetic Particle
  Events}.
\newblock \emph{Astrophys. J. Lett.} 571, L63--L66.
\newblock \doi{10.1086/341149}
\bibAnnoteFile{2002ApJ...571L..63R}

\bibitem[{{Reames}(2013)}]{2013SSRv..175...53R}
{Reames}, D.~V. (2013).
\newblock {The Two Sources of Solar Energetic Particles}.
\newblock \emph{Space Sci. Rev.} 175, 53--92.
\newblock \doi{10.1007/s11214-013-9958-9}
\bibAnnoteFile{2013SSRv..175...53R}

\bibitem[{{Reames}(2018)}]{2018SSRv..214...61R}
{Reames}, D.~V. (2018).
\newblock {Abundances, Ionization States, Temperatures, and FIP in Solar
  Energetic Particles}.
\newblock \emph{Space Sci. Rev.} 214, 61.
\newblock \doi{10.1007/s11214-018-0495-4}
\bibAnnoteFile{2018SSRv..214...61R}

\bibitem[{{Reames}(2020)}]{2020SSRv..216...20R}
{Reames}, D.~V. (2020).
\newblock {Four Distinct Pathways to the Element Abundances in Solar Energetic
  Particles}.
\newblock \emph{Space Sci. Rev.} 216, 20.
\newblock \doi{10.1007/s11214-020-0643-5}
\bibAnnoteFile{2020SSRv..216...20R}

\bibitem[{{Reames}(2021{\natexlab{a}})}]{10.3389/fspas.2021.760261}
{Reames}, D.~V. (2021{\natexlab{a}}).
\newblock {Fifty Years of $^{3}$He-Rich Events}.
\newblock \emph{Frontiers in Astronomy and Space Sciences} 8, 164.
\newblock \doi{10.3389/fspas.2021.760261}
\bibAnnoteFile{10.3389/fspas.2021.760261}

\bibitem[{{Reames}(2021{\natexlab{b}})}]{2021SSRv..217...72R}
{Reames}, D.~V. (2021{\natexlab{b}}).
\newblock {Sixty Years of Element Abundance Measurements in Solar Energetic
  Particles}.
\newblock \emph{Space Sci. Rev.} 217, 72.
\newblock \doi{10.1007/s11214-021-00845-4}
\bibAnnoteFile{2021SSRv..217...72R}

\bibitem[{{Reames} et~al.(1994){Reames}, {Meyer}, and {von
  Rosenvinge}}]{1994ApJS...90..649R}
{Reames}, D.~V., {Meyer}, J.~P., and {von Rosenvinge}, T.~T. (1994).
\newblock {Energetic-Particle Abundances in Impulsive Solar Flare Events}.
\newblock \emph{Astrophys. J. Suppl. Ser.} 90, 649.
\newblock \doi{10.1086/191887}
\bibAnnoteFile{1994ApJS...90..649R}

\bibitem[{{Reames} and {Stone}(1986)}]{1986ApJ...308..902R}
{Reames}, D.~V. and {Stone}, R.~G. (1986).
\newblock {The Identification of Solar $^{3}$He-rich Events and the Study of
  Particle Acceleration at the Sun}.
\newblock \emph{Astrophys. J.} 308, 902.
\newblock \doi{10.1086/164560}
\bibAnnoteFile{1986ApJ...308..902R}

\bibitem[{{Richardson} et~al.(2014){Richardson}, {von Rosenvinge}, {Cane},
  {Christian}, {Cohen}, {Labrador} et~al.}]{2014SoPh..289.3059R}
{Richardson}, I.~G., {von Rosenvinge}, T.~T., {Cane}, H.~V., {Christian},
  E.~R., {Cohen}, C.~M.~S., {Labrador}, A.~W., et~al. (2014).
\newblock {$>$25 MeV Proton Events Observed by the High Energy Telescopes on
  the STEREO A and B Spacecraft and/or at Earth During the First
  {\ensuremath{\sim}} Seven Years of the STEREO Mission}.
\newblock \emph{Sol. Phys.} 289, 3059--3107.
\newblock \doi{10.1007/s11207-014-0524-8}
\bibAnnoteFile{2014SoPh..289.3059R}

\bibitem[{{Rouillard} et~al.(2012){Rouillard}, {Sheeley}, {Tylka}, {Vourlidas},
  {Ng}, {Rakowski} et~al.}]{2012ApJ...752...44R}
{Rouillard}, A.~P., {Sheeley}, N.~R., {Tylka}, A., {Vourlidas}, A., {Ng},
  C.~K., {Rakowski}, C., et~al. (2012).
\newblock {The Longitudinal Properties of a Solar Energetic Particle Event
  Investigated Using Modern Solar Imaging}.
\newblock \emph{Astrophys. J.} 752, 44.
\newblock \doi{10.1088/0004-637X/752/1/44}
\bibAnnoteFile{2012ApJ...752...44R}

\bibitem[{{Shen} et~al.(2018{\natexlab{a}}){Shen}, {Liu}, {Liu}, {Su}, {Tang},
  and {Miao}}]{2018ApJ...861..105S}
{Shen}, Y., {Liu}, Y., {Liu}, Y.~D., {Su}, J., {Tang}, Z., and {Miao}, Y.
  (2018{\natexlab{a}}).
\newblock {Homologous Large-amplitude Nonlinear Fast-mode Magnetosonic Waves
  Driven by Recurrent Coronal Jets}.
\newblock \emph{Astrophys. J.} 861, 105.
\newblock \doi{10.3847/1538-4357/aac9be}
\bibAnnoteFile{2018ApJ...861..105S}

\bibitem[{{Shen} et~al.(2018{\natexlab{b}}){Shen}, {Tang}, {Miao}, {Su}, and
  {Liu}}]{2018ApJ...860L...8S}
{Shen}, Y., {Tang}, Z., {Miao}, Y., {Su}, J., and {Liu}, Y.
  (2018{\natexlab{b}}).
\newblock {EUV Waves Driven by the Sudden Expansion of Transequatorial Loops
  Caused by Coronal Jets}.
\newblock \emph{Astrophys. J. Lett.} 860, L8.
\newblock \doi{10.3847/2041-8213/aac8dd}
\bibAnnoteFile{2018ApJ...860L...8S}

\bibitem[{{Temerin} and {Roth}(1992)}]{1992ApJ...391L.105T}
{Temerin}, M. and {Roth}, I. (1992).
\newblock {The Production of $^{3}$He and Heavy Ion Enrichments in
  $^{3}$He-rich Flares by Electromagnetic Hydrogen Cyclotron Waves}.
\newblock \emph{Astrophys. J. Lett.} 391, L105.
\newblock \doi{10.1086/186408}
\bibAnnoteFile{1992ApJ...391L.105T}

\bibitem[{{Thompson} and {Myers}(2009)}]{2009ApJS..183..225T}
{Thompson}, B.~J. and {Myers}, D.~C. (2009).
\newblock {A Catalog of Coronal ``EIT Wave'' Transients}.
\newblock \emph{Astrophys. J. Suppl. Ser.} 183, 225--243.
\newblock \doi{10.1088/0067-0049/183/2/225}
\bibAnnoteFile{2009ApJS..183..225T}

\bibitem[{{Torsti} et~al.(1999){Torsti}, {Kocharov}, {Teittinen}, and
  {Thompson}}]{1999ApJ...510..460T}
{Torsti}, J., {Kocharov}, L.~G., {Teittinen}, M., and {Thompson}, B.~J. (1999).
\newblock {Injection of $\gtrsim$10 MeV Protons in Association with a Coronal
  Moreton Wave}.
\newblock \emph{Astrophys. J.} 510, 460--465.
\newblock \doi{10.1086/306581}
\bibAnnoteFile{1999ApJ...510..460T}

\bibitem[{{Wang} et~al.(2016){Wang}, {Krucker}, {Mason}, {Lin}, and
  {Li}}]{2016A&A...585A.119W}
{Wang}, L., {Krucker}, S., {Mason}, G.~M., {Lin}, R.~P., and {Li}, G. (2016).
\newblock {The injection of ten electron/$^{3}$He-rich SEP events}.
\newblock \emph{Astron. Astrophys.} 585, A119.
\newblock \doi{10.1051/0004-6361/201527270}
\bibAnnoteFile{2016A&A...585A.119W}

\bibitem[{{Wang} et~al.(2012){Wang}, {Lin}, {Krucker}, and
  {Mason}}]{2012ApJ...759...69W}
{Wang}, L., {Lin}, R.~P., {Krucker}, S., and {Mason}, G.~M. (2012).
\newblock {A Statistical Study of Solar Electron Events over One Solar Cycle}.
\newblock \emph{Astrophys. J.} 759, 69.
\newblock \doi{10.1088/0004-637X/759/1/69}
\bibAnnoteFile{2012ApJ...759...69W}

\bibitem[{{Wang}(2000)}]{2000ApJ...543L..89W}
{Wang}, Y.~M. (2000).
\newblock {EIT Waves and Fast-Mode Propagation in the Solar Corona}.
\newblock \emph{Astrophys. J. Lett.} 543, L89--L93.
\newblock \doi{10.1086/318178}
\bibAnnoteFile{2000ApJ...543L..89W}

\bibitem[{{Wang} et~al.(2006){Wang}, {Pick}, and {Mason}}]{2006ApJ...639..495W}
{Wang}, Y.~M., {Pick}, M., and {Mason}, G.~M. (2006).
\newblock {Coronal Holes, Jets, and the Origin of $^{3}$He-rich Particle
  Events}.
\newblock \emph{Astrophys. J.} 639, 495--509.
\newblock \doi{10.1086/499355}
\bibAnnoteFile{2006ApJ...639..495W}

\bibitem[{{Warmuth}(2015)}]{2015LRSP...12....3W}
{Warmuth}, A. (2015).
\newblock {Large-scale Globally Propagating Coronal Waves}.
\newblock \emph{Liv. Rev. Sol. Phys.} 12, 3.
\newblock \doi{10.1007/lrsp-2015-3}
\bibAnnoteFile{2015LRSP...12....3W}

\bibitem[{{Warmuth} and {Mann}(2011)}]{2011A&A...532A.151W}
{Warmuth}, A. and {Mann}, G. (2011).
\newblock {Kinematical evidence for physically different classes of large-scale
  coronal EUV waves}.
\newblock \emph{Astron. Astrophys.} 532, A151.
\newblock \doi{10.1051/0004-6361/201116685}
\bibAnnoteFile{2011A&A...532A.151W}

\bibitem[{{Wiedenbeck} et~al.(2020){Wiedenbeck}, {Bu{\v{c}}{\'\i}k}, {Mason},
  {Ho}, {Leske}, {Cohen} et~al.}]{2020ApJS..246...42W}
{Wiedenbeck}, M.~E., {Bu{\v{c}}{\'\i}k}, R., {Mason}, G.~M., {Ho}, G.~C.,
  {Leske}, R.~A., {Cohen}, C.~M.~S., et~al. (2020).
\newblock {$^{3}$He-rich Solar Energetic Particle Observations at the Parker
  Solar Probe and near Earth}.
\newblock \emph{Astrophys. J. Suppl. Ser.} 246, 42.
\newblock \doi{10.3847/1538-4365/ab5963}
\bibAnnoteFile{2020ApJS..246...42W}

\bibitem[{{Wiedenbeck} et~al.(2013){Wiedenbeck}, {Mason}, {Cohen}, {Nitta},
  {G{\'o}mez-Herrero}, and {Haggerty}}]{2013ApJ...762...54W}
{Wiedenbeck}, M.~E., {Mason}, G.~M., {Cohen}, C.~M.~S., {Nitta}, N.~V.,
  {G{\'o}mez-Herrero}, R., and {Haggerty}, D.~K. (2013).
\newblock {Observations of Solar Energetic Particles from $^{3}$He-rich Events
  over a Wide Range of Heliographic Longitude}.
\newblock \emph{Astrophys. J.} 762, 54.
\newblock \doi{10.1088/0004-637X/762/1/54}
\bibAnnoteFile{2013ApJ...762...54W}

\bibitem[{{Zhang} and {Zhao}(2017)}]{2017ApJ...846..107Z}
{Zhang}, M. and {Zhao}, L. (2017).
\newblock {Precipitation and Release of Solar Energetic Particles from the
  Solar Coronal Magnetic Field}.
\newblock \emph{Astrophys. J.} 846, 107.
\newblock \doi{10.3847/1538-4357/aa86a8}
\bibAnnoteFile{2017ApJ...846..107Z}

\bibitem[{{Zheng} et~al.(2012{\natexlab{a}}){Zheng}, {Jiang}, {Yang}, {Bi},
  {Hong}, {Yang} et~al.}]{2012A&A...541A..49Z}
{Zheng}, R., {Jiang}, Y., {Yang}, J., {Bi}, Y., {Hong}, J., {Yang}, B., et~al.
  (2012{\natexlab{a}}).
\newblock {An extreme ultraviolet wave associated with a failed eruption
  observed by the Solar Dynamics Observatory}.
\newblock \emph{Astron. Astrophys.} 541, A49.
\newblock \doi{10.1051/0004-6361/201118305}
\bibAnnoteFile{2012A&A...541A..49Z}

\bibitem[{{Zheng} et~al.(2012{\natexlab{b}}){Zheng}, {Jiang}, {Yang}, {Bi},
  {Hong}, {Yang} et~al.}]{2012ApJ...753..112Z}
{Zheng}, R., {Jiang}, Y., {Yang}, J., {Bi}, Y., {Hong}, J., {Yang}, D., et~al.
  (2012{\natexlab{b}}).
\newblock {A Fast Propagating Extreme-Ultraviolet Wave Associated with a
  Mini-filament Eruption}.
\newblock \emph{Astrophys. J.} 753, 112.
\newblock \doi{10.1088/0004-637X/753/2/112}
\bibAnnoteFile{2012ApJ...753..112Z}

\end{thebibliography}

\end{document}